\documentclass[]{spie}  

\usepackage{amsmath,amsfonts,amssymb}
\usepackage{graphicx}
\usepackage[colorlinks=true, allcolors=blue]{hyperref}

\title{Broadband vortex fiber nulling: high-dispersion exoplanet science at the diffraction limit}

\author[a,$\dagger$]{Daniel~Echeverri}
\author[b]{Garreth~Ruane}
\author[a]{Nemanja~Jovanovic}
\author[c]{Jacques-Robert~Delorme}
\author[a]{Jason~Wang}
\author[d]{Maxwell~A.~Millar-Blanchaer}
\author[a]{Jerry~Xuan}
\author[a]{Katie Toman}
\author[a,b]{Dimitri~Mawet}
\affil[a]{Department of Astronomy, California Institute of Technology, 1200 E. California Blvd.\\Pasadena, CA 91125, USA}
\affil[b]{Jet Propulsion Laboratory, California Institute of Technology, 4800 Oak Grove Dr.
\\Pasadena, CA 91109, USA}
\affil[c]{W. M. Keck Observatory, 65-1120 Mamalahoa Highway, Kamuela, HI 96743, USA.}
\affil[d]{Department of Physics, University of California, Santa Barbara, CA 93106, USA}

\authorinfo{$^\dagger$Send correspondence to dechever@caltech.edu}

\graphicspath{{Figures/}}
\begin{document} 
\maketitle

\begin{abstract}
As the number of confirmed exoplanets continues to grow, there is an increased push to spectrally characterize them to determine their atmospheric composition, formation paths, rotation rates, and habitability. However, there is a large population of known exoplanets that either do not transit their star or have been detected via the radial velocity (RV) method at very small angular separations such that they are inaccessible to traditional coronagraph systems. Vortex Fiber Nulling (VFN) is a new single-aperture interferometric technique that uses the entire telescope pupil to bridge the gap between traditional coronagraphy and RV or Transit methods by enabling the direct observation and spectral characterization of targets at and within the diffraction limit. By combining a vortex mask with a single mode fiber, the on-axis starlight is rejected while the off-axis planet light is coupled and efficiently routed to a radiometer or spectrograph for analysis. We have demonstrated VFN in the lab monochromatically in the past. In this paper we present a polychromatic validation of VFN with nulls of $<10^{-4}$ across 15\% bandwidth light. We also provide an update on deployment plans and predicted yield estimates for the VFN mode of the Keck Planet Imager and Characterizer (KPIC) instrument. Using PSISIM, a simulation package developed in cooperation with several groups, we assess KPIC VFN’s ability to detect and characterize different types of targets including planet candidates around promising young-moving-group stars as well as known exoplanets detected via the RV method. The KPIC VFN on-sky demonstration will pave the road to deployment on future instruments such as Keck-HISPEC and TMT-MODHIS where it could provide high-resolution spectra of sub-Jupiter mass planets down to 5 milliarcseconds from their star.
\end{abstract}


\section{INTRODUCTION}
\label{sec:intro}  
With the NASA Exoplanet Archive\cite{Akenson2013_Archive} boasting more than 4,400 confirmed exoplanets, there is a growing push to go beyond simple detection and towards characterization. A promising new direction for this characterization endeavour combines high-contrast imaging with high resolution spectroscopy to improve the sensitivity of direct imaging observations\cite{Snellen2015_HDC}. Obtaining high resolution spectra also enables measurements of, among other things, a planet's atmospheric composition via molecular mapping\cite{Konopacky2013_MolecularMapping}, its spin via rotational broadening\cite{Snellen2014_Spin}, and its cloud coverage and weather via Doppler imaging\cite{Crossfield2014_DopplerImagingDemo}. However conventional coronagraphs have an inner working angle (IWA) of $\sim2\lambda/D$, where $\lambda$ is the central wavelength and $D$ is the telescope diameter, which limits the exoplanets that can be observed such that there is a conspicuous gap between transit spectroscopy and direct imaging. Nulling interferometry, as introduced by Bracewell in 1978 \cite{Bracewell1978}, can beat the typical IWA and target exoplanets at smaller separations, thereby expanding the available search space for exoplanet detection and characterization campaigns to bridge the gap between transit and direct imaging.

Nulling interferometry has taken several forms in the past, even being considered for major space telescope missions such as NASA's Terrestrial Planet Finder Interferometer\cite{Lawson2007_TPFI} and ESA's Darwin\cite{Kaltenegger2005_Darwin}. In more recent years, versions relying on optical fibers to simplify the beam combination, known as Fiber Nulling, have arisen. The Palomar Fiber Nuller (PFN)\cite{Haguenauer2006_PFN,Mennesson2011_PFN} is an example of such an instrument which has demonstrated on-sky starlight suppression on the order of $10^{-4}$ with the detection of a faint companion at a separation of 30~mas, well within the 88~mas diffraction limit of the Palomar Telescope in K band\cite{Serabyn2019_PFN}. Building on that heritage, our team at Caltech has developed and experimentally demonstrated a new fiber nulling technique, known as Vortex Fiber Nulling (VFN).

In this paper, we review the VFN concept and previous laboratory demonstrations. We then present new results that validate the VFN technique in broadband light. We demonstrate deep starlight suppression down to $<10^{-4}$ at 15\% bandwidth using commercially available, off-the-shelf optics and we confirm that the system is well understood by comparing the lab results with our system model. Finally, we present yield estimates for the planned VFN mode of the Keck Planet Imager and Characterizer (KPIC) instrument at the Keck II Telescope. We find that KPIC VFN, though primarily an on-sky demonstrator of the VFN concept, can detect giant planets down to a few Jupiter masses at unprecedented small angular separations that would be unreachable with current conventional coronagraphs

\section{VFN CONCEPT}
\label{sec:VFN_Concept}

\begin{figure}[t!]
    \centering
    \includegraphics[width=0.85\linewidth]{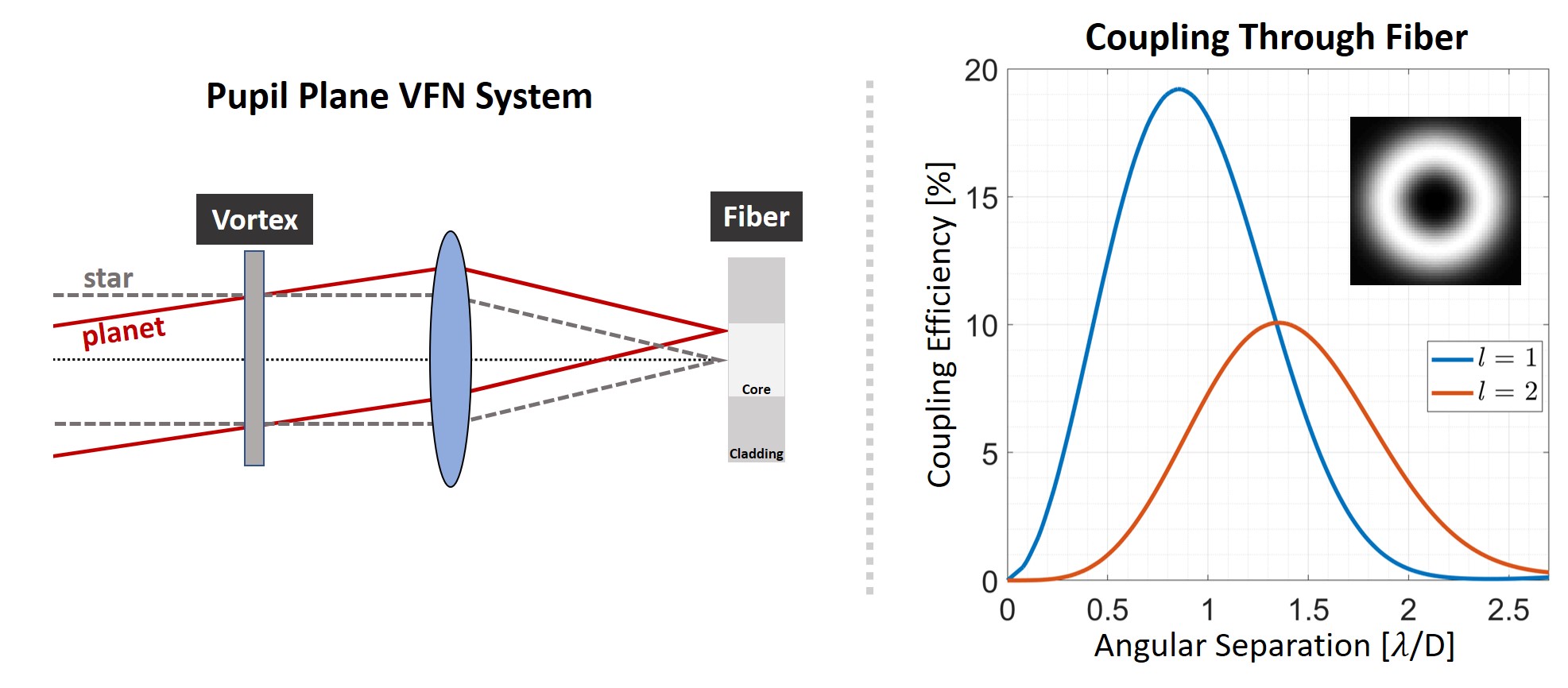}
    \caption{(Left) Schematic of a VFN system with the vortex mask in a pupil plane before a single mode fiber. For VFN, the optical fiber is aligned with the on-axis star such that the planet lands off-center on the fiber. (Right) Coupling profile for VFN. The blue and orange curves show the coupling efficiency as a fraction of light incident on the fiber tip for charge 1 and 2 vortex masks respectively. These coupling curves are azimuthally symmetric as shown in the inset image.}
    \label{fig:MotherFig}
\end{figure}

Vortex Fiber Nulling is a single-aperture, interferometric technique for suppressing starlight while simultaneously transmitting the light from an off-axis companion for detection and characterization\cite{Ruane2018_VFN,Ruane2019SPIE,Echeverri2019_VFN}. Like previous fiber nulling techniques, VFN leverages the modal filtering properties of a single mode fiber (SMF) to achieve the nulling effect but instead of using an aperture mask to pick off parts of the pupil, it interferes the full aperture with itself. This is achieved using an optical vortex mask that imparts a phase ramp of the form $\exp(il\theta)$ where $l$ is an integer, known as the charge, which represents the number of times the phase wraps from 0 to 2$\pi$ and $\theta$ is the azimuthal coordinate angle. The vortex can be placed anywhere along the optical path but for this paper, we focus on the pupil plane version of VFN as shown in the schematic on the left in Fig.~\ref{fig:MotherFig}.

During an observation, the SMF is aligned with the on-axis star such that the star's vortex phase is symmetric over the fundamental mode of the fiber. This makes the two orthogonal and the starlight does not couple into the fiber. This also puts the light from the planet off-axis on the fiber and thus makes it not symmetric on the fiber mode such that it couples in with the angle-dependent efficiency shown in the right plot from Fig.~\ref{fig:MotherFig}. The peak planet coupling efficiency, defined as the maximum coupling for an off-axis point, is 19\% with charge 1 for a planet at 0.9$\lambda/D$ and 10\% with charge 2 at 1.4$\lambda/D$. Charge 1 still gets more than 10\% coupling efficiency for a planet at 0.4$\lambda/D$ while charge 2 maintains more than 5\% coupling out to a separation of 1.9$\lambda/D$. This puts the entire field of view of VFN within the inner working angle of conventional coronagraphs. 

VFN has several benefits that are presented in more detail in Ruane et al. 2019\cite{Ruane2019SPIE}. The first, key, benefit is that the coupling efficiency is azimuthally symmetric as shown in the inset on the right of Fig.~\ref{fig:MotherFig}. As such, knowledge of the exact planet position is not needed for an observation; as long as the planet lands somewhere in the single, radial bright VFN fringe, it will couple into the fiber. This symmetry also removes the need for baseline rotation required by other nulling techniques and the planet can be spectrally detected rather than radiometrically. Other benefits are that VFN is insensitive to the pupil shape, has a large null space of Zernike modes, and can be readily implemented on any high contrast imaging system with a fiber injection unit\cite{Ruane2019SPIE}. Once the light is in the fiber, it can be easily and efficiently routed to a high-resolution spectrograph where post-processing techniques can further disentangle the star and planet light and enable the detection and detailed characterization of the exoplanet\cite{Wang2017,Snellen2015_HDC}.

\section{LABORATORY DEMONSTRATION}
\label{sec:Lab_Demos}

\subsection{Laboratory Setup}
\label{sec:lab_setup}

\begin{figure}[t!]
    \centering
    \includegraphics[width=0.9\linewidth]{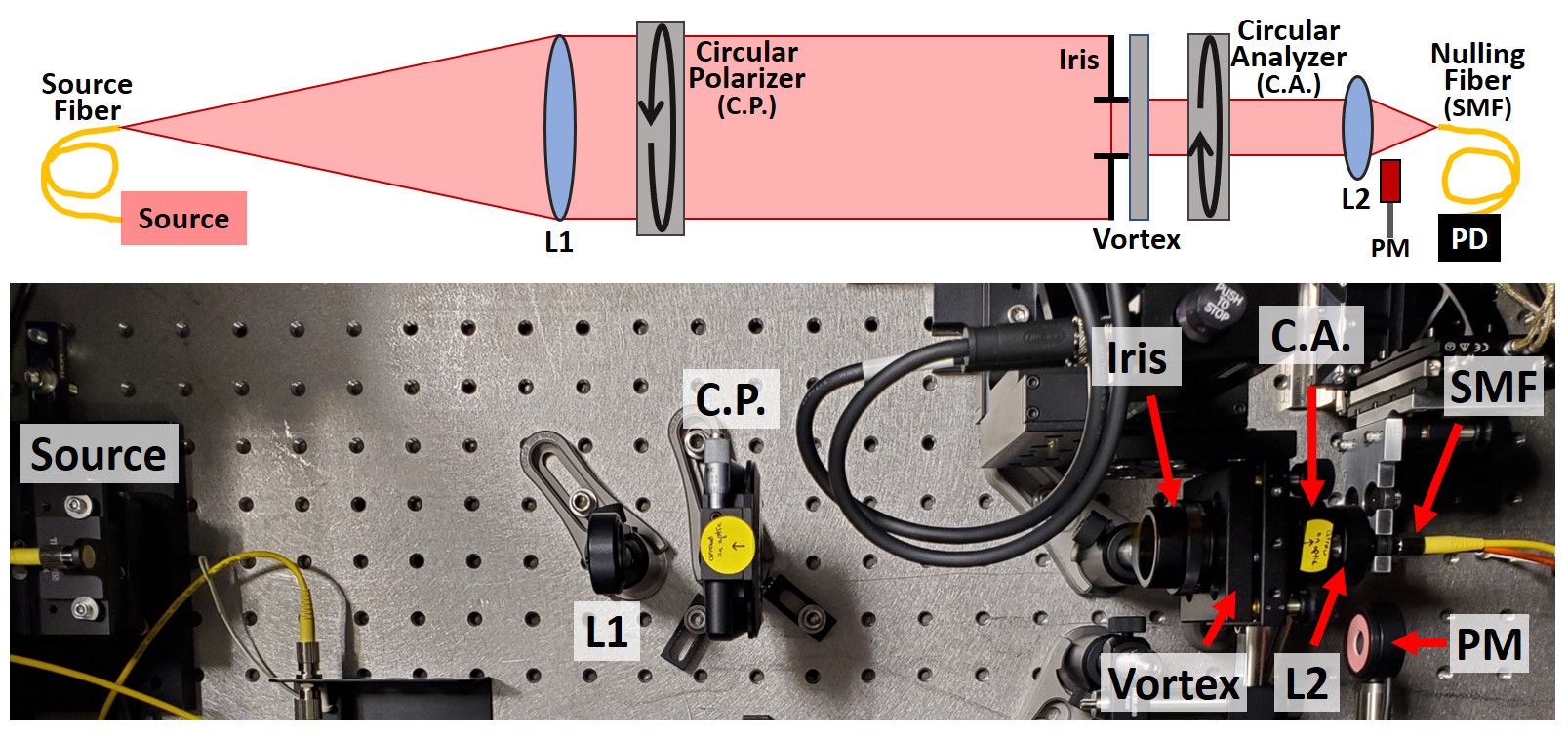}
    \caption{(Top) Schematic of the VFN testbed at Caltech. Light from a single mode fiber (SMF) source is collimated by lens L1 before passing sequentially through a circular polarizer (C.P.), iris, vortex mask, and circular analyzer (C.A.). Lens L2 then focuses the light onto an SMF that is connected to a photodiode (PD) which measures the power through the fiber. An optical power meter (PM) can be translated into the optical path between L2 and the second SMF to measure the power incident on the fiber. (Bottom) Matching picture of the testbed as described by the schematic.}
    \label{fig:VFNBench}
\end{figure}

We tested the VFN concept in the VFN testbed at the Caltech Exoplanet Technologies Lab. The testbed, shown in Fig.~\ref{fig:VFNBench}, is relatively simple, reflecting the fact that VFN requires very few optics to achieve the nulling effect. For most of these experiments we used one of two light sources: a fiber-coupled 635~nm laser (Thorlabs~KLS635) for monochromatic tests, and a supercontinuum white light source (NKT Photonics SuperK EXTREME) with a tunable filter (NKT Photonics SuperK VARIA) for broadband tests. The source was injected into the testbed through an SMF and the resulting light was collimated by lens L1 before passing through a circular polarizer (C.P.) and iris. We used a charge 2 polychromatic vector vortex mask in the broadband tests presented below though we have also used a monochromatic charge 1 vector vortex mask in the past. We also used a circular analyzer to filter the residual chromatic leakage imparted by our imperfect vector vortex mask and ensure that only a single polarization of light was left in the system (see Llop-Sayson et al., these proceedings) before focusing the light onto the tip of another SMF which acted as the nulling fiber. The power through that fiber was measured with a variable-gain photodiode (PD - Femto OE-200-SI) while the power incident on the fiber was measured with a separate power meter (PM - Thorlabs PM100D). Normalizing the power through the fiber by the power incident on the fiber provided the coupling efficiency of the system. We translated the nulling fiber in the focal plane using high precision actuators (PI Q-545.240) which allowed us to measure the stellar rejection, referred to here as the null depth, for the on-axis source along with the coupling efficiency at various off-axis separations.

We used this same testbed before to demonstrate excellent VFN performance in monochromatic light including null depths of about $5\times10^{-5}$ with both charge 1 and charge 2 vortex masks\cite{Echeverri2019_VFN,Echeverri2020_VFNSPIE}. We didn't use the circular polarizer and analyzer for those experiments since, in monochromatic light, we have enough degrees of freedom in moving the vortex and SMF to achieve deep nulls regardless of leakage from the vortex mask. However, for the broadband experiments we found that we were limited by chromatic leakage from the vector vortex mask so adding the polarizers let us work in a single circular polarization state as described in Sec.~\ref{sec:broadband_vali}.

\subsection{Broadband Validation}
\label{sec:broadband_vali}

\begin{figure}[t!]
    \centering
    \includegraphics[width=0.5\linewidth]{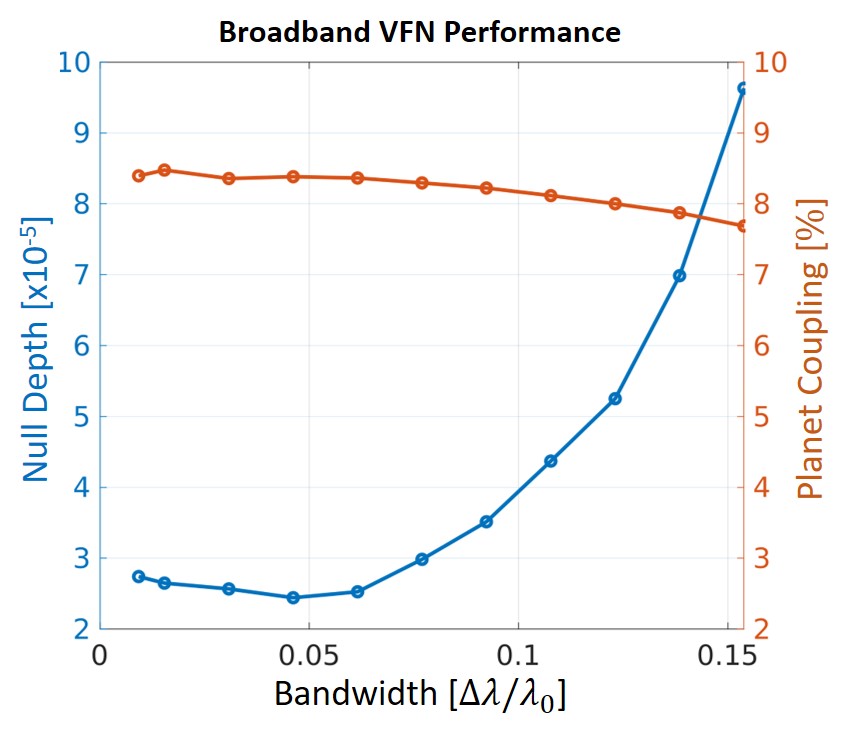}
    \caption{Broadband performance of VFN in the laboratory. The blue curve shows the null depth and corresponds to the left y-axis while the orange curve is the peak planet coupling and corresponds to the right y-axis. These experiments were performed at various bandwidths centered around 650~nm light such that a fractional bandwidth of 0.15 corresponds to a bandpass from about 600 to 700~nm.}
    \label{fig:VFNBroadbandResults}
\end{figure}

We previously reported null depths of $4.5\times10^{-4}$ with simultaneous peak planet coupling efficiencies of 4.5\% in 10\% bandwidth light\cite{Echeverri2019b_VFN,Echeverri2020_VFNSPIE}. Those results were centered around 790~nm light since that produced the deepest nulls. However, we found at the time that based on the amount of wavefront error in our system, we should have been able to achieve better null depths. This pointed to chromatic leakage in the vector vortex mask being the limiting factor in the system performance, not wavefront errors. A practical flaw in vector vortex masks is that there is a wavelength-dependent retardance error which results in a fraction of the incoming light being transmitted without obtaining the azimuthal vortex phase\cite{Ruane2019_scalarVC}. This leaked light couples into the SMF and results in stellar leakage due to errors in the vortex mask itself. Though we don't have measurements of the retardance error for our current charge 2 mask, it seems we were leakage-limited at about the $4\times10^{-4}$ level with the leakage getting worse at shorter wavelengths. 

For a given circular polarization state, right-handed or left-handed, the leakage term from a vector vortex mask appears as similarly-polarized light after the vortex while the light with the desired phase ramp switches to the opposite polarization state\cite{Ruane2019_scalarVC}. We can leverage this by working with a single polarization state so that the chromatic leakage is rejected from the system. This is achieved by adding opposite-handed polarizers up and downstream of the vortex mask. The upstream polarizer, labeled C.P. in Fig.~\ref{fig:VFNBench}, prepares the light so that the vortex only sees a single polarization, right-hand circular in this case. The vortex imparts its azimuthal phase ramp and flips the polarization to left-hand circular. We can then filter for left-hand circular light using the circular analyzer (i.e. the second circular polarizer) so that the leakage, which is right-handed in our system, is removed and we can achieve much deeper nulls. 

With the circular polarizer (Thorlabs CP1R633) and analyzer (Thorlabs CP1L633) in the optical path, the system performance improved by an order of magnitude and we were able to switch to shorter wavelengths. As shown in blue in Fig.~\ref{fig:VFNBroadbandResults}, we achieved a null of $4.0\times10^{-5}$ in 10\% bandwidth light centered around 650~nm. Even at 15\% bandwidth the null remained under $10^{-4}$ which is a factor of 5 better than previous results at 10\% bandwidth without the polarizers. The null plateaus at around $2.5\times10^{-5}$ for bandwidths less than about 7\%. This plateau shows a transition from the wavefront-limited to leakage-limited regime. The null of $2.5\times10^{-5}$ is a very good match to our system model which predicts nulls of $1.4\times10^{-5}$ based primarily on the measured wavefront error in the system. The null starts to degrade more rapidly beyond 7\% bandwidth which is most likely due to the fact that our polarizers were designed for monochromatic 633~nm light. This means that the chromatic leakage from the vortex is poorly filtered towards the edges of the band and starts to dominate the null performance for larger bandwidths. With a better polychromatic vector vortex mask, we would no longer be limited by the chromatic leakage and could remove the polarizers from the system entirely. 

In addition to the polarizers, we also modified the system from previous experiments to optimize the system F/\#. This was done by manually tuning the iris diameter so that we could set the pupil size that could maximize the coupling into the fiber. The orange curve in Fig.~\ref{fig:VFNBroadbandResults} displays the resulting peak planet coupling at each bandwidth. We achieved a peak coupling of about 8.1\% at 10\% bandwidth which is almost a factor of 2 better than we had done before. This coupling is close to the theoretical maximum of 10\% for a charge 2 vortex. We are now tracking down what un-modeled losses could be causing the small discrepancy in peak coupling. However, given the nice agreement with the predicted null-depth and the close-to-ideal planet coupling, these results demonstrate that VFN works in broadband light.


\subsection{Next Steps In-Lab}
\label{sec:lab_next_step}
With the broadband results presented above, we have experimentally demonstrated that VFN works in broadband light and that we understand the performance we are achieving in the lab. The next step is to build the Polychromatic Reflective Testbed (PoRT) as described in a previous proceedings paper\cite{Echeverri2020_VFNSPIE}. PoRT provides several improvements over the current testbed including that it uses higher-quality reflective optics that will make the system performance less wavelength-dependent, allowing us to use the same system for both visible and near-infrared experiments.  We will first repeat the visible broadband experiments on PoRT to validate the system before quickly moving on to K-band (2.0-2.4~um) experiments. These K-band tests will verify that VFN works as expected at longer wavelengths and will allow us to test and pick the vortex mask that will be deployed to the KPIC instrument next year. We currently have two charge 2 vector vortex masks that we can choose from so by testing both in the lab, we can determine which one has the best performance and use that for our on-sky demonstration of VFN. Some of these experiments are already under way and were detailed in last year's paper\cite{Echeverri2020_VFNSPIE}.

Once we have chosen a mask for KPIC, we will move the K-band experiments to the KPIC Phase II plate, which will be described in the next section, Sec.~\ref{sec:KPIC_VFN_intro}. We will then expand PoRT to add a deformable mirror (DM) which we will use to start testing the sensitivity of VFN to different wavefront aberrations explicitly and individually in the lab. We previously reported simulations showing that VFN is sensitive to specific aberrations based on the vortex charge that is used and we provided curves that show how the null degrades with increasing amplitudes of aberration\cite{Ruane2019SPIE}. These in-lab experiments will serve to validate those curves. More specifically, we'd like to verify that charge 1 is primarily sensitive to coma while charge 2 is sensitive to astigmatism.

\section{KPIC VFN YIELD ESTIMATES}
\label{sec:KPIC_VFN_intro}
With broadband VFN validated in the lab, we are preparing for an on-sky demonstration with the Keck Planet Imager and Characterizer (KPIC) instrument\cite{Mawet2016_KPIC,Delorme2021_KPIC,Wang2021_KPICScience}. KPIC is an upgrade to the Keck II Adaptive Optics (AO) system that adds a near-infrared pyramid wavefont sensor (PyWFS)\cite{Bond2018_PyWFS} as well as a fiber injection unit (FIU) connected to \mbox{NIRSPEC}\cite{Martin2018}, the extant near-infrared spectrograph at the Keck II Telescope. KPIC thus provides AO-corrected, high-resolution spectra with a resolving power of 35,000 covering J-H-K bands (1.1-2.4~um). 

KPIC is a phased upgrade meaning that it has several versions with each phase being deployed and commissioned individually before moving on to the next one. We finished commissioning Phase I in 2020\cite{Delorme2021_KPIC} and have obtained high resolution spectra of several targets including the HR-8799 planets. This allowed the KPIC team to determine the effective temperatures, surface gravities, radial velocities, and spins of HR-8799 c, d, and e\cite{Wang2021_KPICScience}. 

Phase II is now being assembled and tested at Caltech for deployment in early 2022. This phase brings several upgrades to the instrument including the first on-sky demonstration of VFN, described in detail in our proceedings paper from 2019\cite{Echeverri2019b_VFN}. The main additions enabling the KPIC VFN mode are the high-order DM and vortex mask. We will initially deploy a charge 2 vector vortex mask since this relaxes the tip-tilt (TT) jitter requirements which may be the dominant source of null depth degradation early in KPIC VFN deployment. However, we are looking to either swap to, or add, a charge 1 vector vortex mask if the system performance allows it. This is because our simulations, below and in previous papers\cite{Echeverri2019b_VFN}, show that the increased throughput and smaller inner-working angle provided by charge 1 VFN increase the scientific yield significantly. 

As we prepare for KPIC VFN, we are starting to gather a preliminary target list along with running simulations to refine the VFN science case on the 10~m aperture of the Keck Telescope. These simulations not only provide motivation for the on-sky VFN demonstration but they are also helping guide what type of exoplanets we target first. We used a python package called PSISIM for the simulations.

\subsection{PSISIM Simulations Package}
\label{sec:PSISIM}
PSISIM\footnote{\url{https://github.com/planetarysystemsimager/psisim}} is an open-source python package written by a collaboration of institutions (primarily Caltech and the University of California) and is meant to simulate exoplanet observations with a variety of telescopes and instruments. At its core, PSISIM is an exposure time calculator that takes in spectra, simulates an observation, and provides the final spectra which can be processed to determine the cross-correlation function (CCF) signal-to-noise ratio (SNR). The core of PSISIM is based on concepts similar to the simulator described by Wang et al. 2016\cite{Wang2016_ETC}, but has been designed to allow for flexible simulation of a range of different user-definable instruments and telescopes. 

\begin{figure}[t!]
    \centering
    \includegraphics[width=0.9\linewidth]{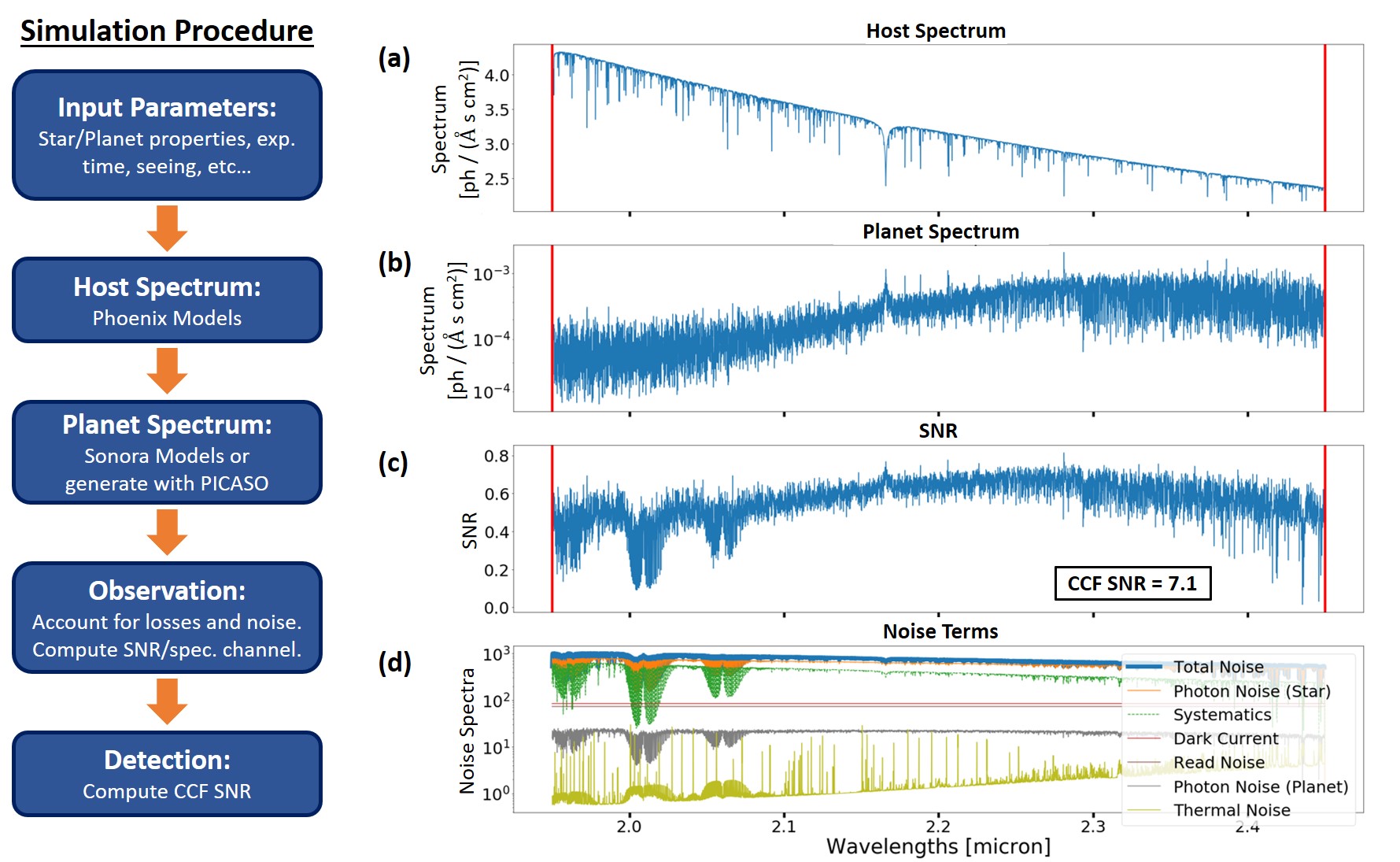}
    \caption{Overview of how a PSISIM simulation is performed. The flowchart on the left details the procedure that is followed for the simulation along with the inputs and considerations that are made at each step. The plots on the right show the spectra considered by PSISIM at various key points during a simulation of Upsilon~Andromeda~b with KPIC VFN. (a) and (b) are the input star and planet spectra respectively. (c) SNR per spectral channel including the noise terms presented in (d). We can compute the CCF~SNR, shown in the box in (c), by cross-correlating the input exoplanet spectrum with the output, noisy spectrum; in this case the simulation computed a CCF~SNR of 7.1.}
    \label{fig:PSISIM_Overview}
\end{figure}

Figure~\ref{fig:PSISIM_Overview} shows a flow-chart of the typical simulation procedure followed by PSISIM along with the spectra at key points during the simulation. An observation of Upsilon~Andromeda~b\cite{Butler1997_UpsAndb} with KPIC VFN is shown as an example. The user sets the observation parameters which includes the star and planet properties, telescope and instrument to use, instrument mode, exposure time, and observing conditions. Based on the stellar properties, PSISIM pulls the matching star spectrum, shown in Fig.~\ref{fig:PSISIM_Overview}(a), from a library of high-resolution PHOENIX models\cite{Husser2013_PHOENIX}. There are two options that PSISIM can use for the exoplanet spectra: (1) it can pull the closest matching spectrum from a grid of SONORA models\cite{Marley2021_SONORA} or (2) it can generate a custom spectrum using PICASO, an open source radiative transfer python package\cite{Batalha2019_PICASO}. Figure~\ref{fig:PSISIM_Overview}(b) shows the spectrum used for Ups~And~b in this example. PSISIM then simulates the observation by accounting for losses and noise terms including, among other effects, atmospheric transmission, telescope and instrument throughput, thermal background, photon noise, and detector noise. It applies these effects to the input spectra to provide the final spectrum as extracted from the science detector as well as the SNR per spectral channel, shown in Fig.~\ref{fig:PSISIM_Overview}(c). The final spectrum can be cross-correlated with the input planet model to compute the CCF~SNR which tells us how detectable the original planet spectrum is over the star's spectrum and the noise present in the observation\cite{Snellen2015_HDC}. This CCF SNR technique is the same method currently used for detecting exoplanets with KPIC\cite{Wang2021_KPICScience}. For Ups~And~b, this simulation predicts that KPIC VFN would achieve a CCF~SNR of about 7.1 in an hour of observation with a charge 1 vortex mask under the conditions applied for the example. A CCF~SNR greater than 5 is considered a detection in this paper. We can also extract the individual noise terms, as shown in Fig.~\ref{fig:PSISIM_Overview}(d), from PSISIM to see what is the limiting effect for a given observation.

All of the VFN observation simulations below follow this same procedure and there is sample code available in the PSISIM repository that shows how to run a PSISIM VFN simulation\footnote{\url{https://github.com/planetarysystemsimager/psisim/blob/a94deccef0a266480987bb590e52e649a9ed96ea/Tutorials/PSISIM\%20VFN\%20Tutorial.ipynb}}. We initially tested KPIC VFN's ability to detect currently-known exoplanets but found that Ups~And~b was the only one that would be detectable with reasonable observing conditions and within a reasonable amount of time. As such, we considered another possible KPIC VFN campaign targeting young moving group stars. 

\subsection{Young Moving Group Targets}
\label{sec:KPIC_YMG}
Younger stars are generally more favorable for direct imaging surveys since their young age makes putative low-mass companions brighter in the near-infrared. Young stars tend to be found in co-moving groups with similar properties and ages; such associations are known as Young Moving Groups (YMG)\cite{Gagne2018_YMGList}. Here we consider a theoretical KPIC VFN campaign targeting YMG stars generally as well as a realistic campaign guided by promising astrometric signals from previous and current astrometry missions. 

\subsubsection{Blind YMG Survey Yields}
\label{sec:KPIC_YMG_BlindSurvey}
\begin{figure}[t!]
    \centering
    \includegraphics[width=0.6\linewidth]{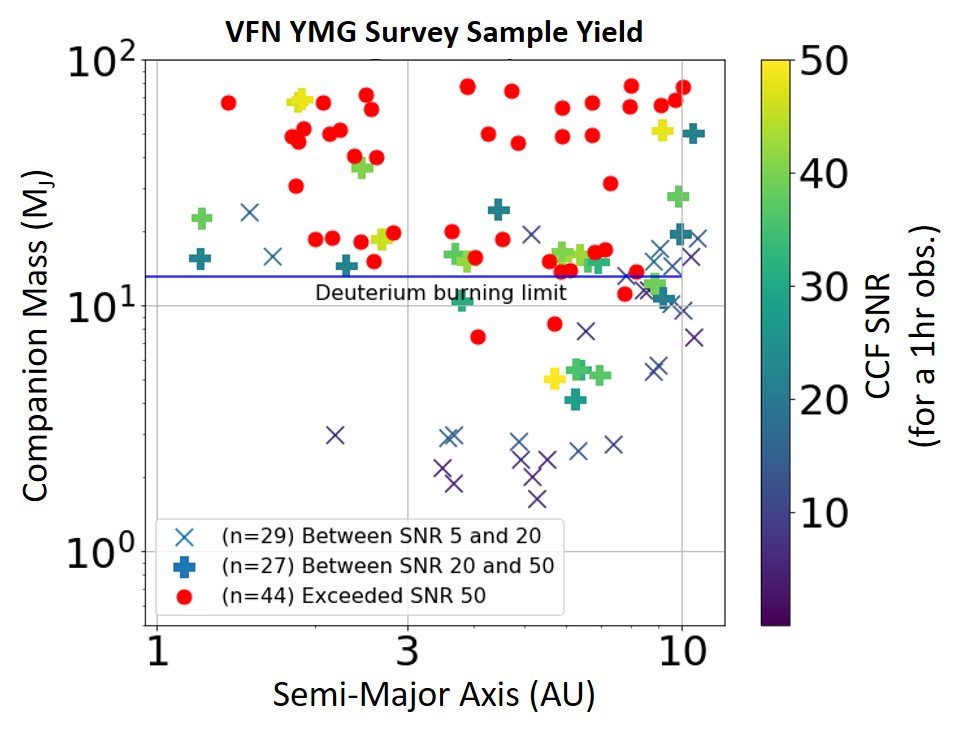}
    \caption{Theoretical yield estimate for a blind KPIC VFN survey of YMG stars. A total of 3,275 stars were placed in our simulated universe and randomly populated with companions based on RV occurrence rates. This figure shows one sample iteration of the random universe. In this particular sample, 635 of the stars have a companion of which 308 are within the field of view for KPIC VFN. Of those 308, 100 companions would be detectable within 1 hour each of observation. This assumes a charge 1 vortex mask with KPIC VFN and the current AO performance measured during Phase I commissioning.}
    \label{fig:PSISIM_YMGYield}
\end{figure}

The first survey we considered is a blind search for hypothetical low-mass companions around YMG stars. It is an unrealistic survey since it would require an unreasonable number of observations but it serves to determine a ball-park estimate of how many exoplanets would be detectable with KPIC VFN in general. Section~\ref{sec:KPIC_YMG_GaidedSurvey} provides a realistic, targeted survey that we are considering for early KPIC VFN science.

We start with a YMG membership list provided by J. Gagn\'e in a private communication and compiled following the same procedure described in Gagn\'e et al. 2018\cite{Gagne2018_YMGList}. We filtered this list for stars visible from the North (declination $>-40^{\circ}$) so that they would be accessible from Keck. This leaves 3,275 stars to consider. We then randomly populated those stars with exoplanets and low-mass companions based on the RV occurrence rates determined by Fulton et al. and Rosenthal et al. 2021\cite{Fulton2021_OccurRates,Rosenthal2021_OccurRates} to create a hypothetical universe of stars and companions. The star properties were set based on the average properties for the associated YMG and the companion properties were determined from the AMES-COND 2000\cite{Allard2001_CONDModels} evolutionary models given the system's age and the companion mass. Finally, we passed these targets through PSISIM to simulate an hour-long observation of each one with a charge 1 mask for VFN on KPIC assuming the current AO performance measured from KPIC Phase I (3~mas TT jitter and approximately 85~nm RMS of low-order wavefront error to which VFN would be sensitive). We repeated the random population step several times to generate multiple iterations of our hypothetical universe so we could sample the resulting yield. 

Figure~\ref{fig:PSISIM_YMGYield} shows a sample iteration of this simulation. In this example, 635 of the 3,275 stars have a companion and 308 of those companions are within VFN field of view. The red points denote companions that would be detected with a CCF~SNR of greater than 50 while the crosses and plus signs denote CCF SNRs greater than 5 and 20 respectively. As expected, the higher-mass objects are more readily detectable since they will be brighter and hence easier to detect behind their host star's glare. However, this plot shows that there are still a few dozen planetary mass companions that could be detected going down to as low as 2 Jupiter masses. Note that these detections are all at separations of between 1 and 10~AU which for typical YMG targets is a region that generally cannot be directly imaged in the near infrared with current telescopes.

Although this hypothetical survey would be able to detect several dozens of low-mass companions, it is very inefficient since the blind nature of it would lead to many non-detections per observable exoplanet. As such, this analysis serves only to motivate and validate the fact that KPIC VFN will provide access to a large population of exoplanets at unprecedented small angular separations that would otherwise be unreachable with current coronagraph technologies. It also shows the power of targeting young stars and searching for planets at the separations where they are most likely to exist. A more practical and manageable KPIC VFN campaign would leverage other exoplanet detection techniques to guide the selection process for its target list. 

\subsubsection{YMG Survey Guided by Gaia and Hipparcos Data}
\label{sec:KPIC_YMG_GaidedSurvey}
\begin{figure}[t!]
    \centering
    \includegraphics[width=\linewidth]{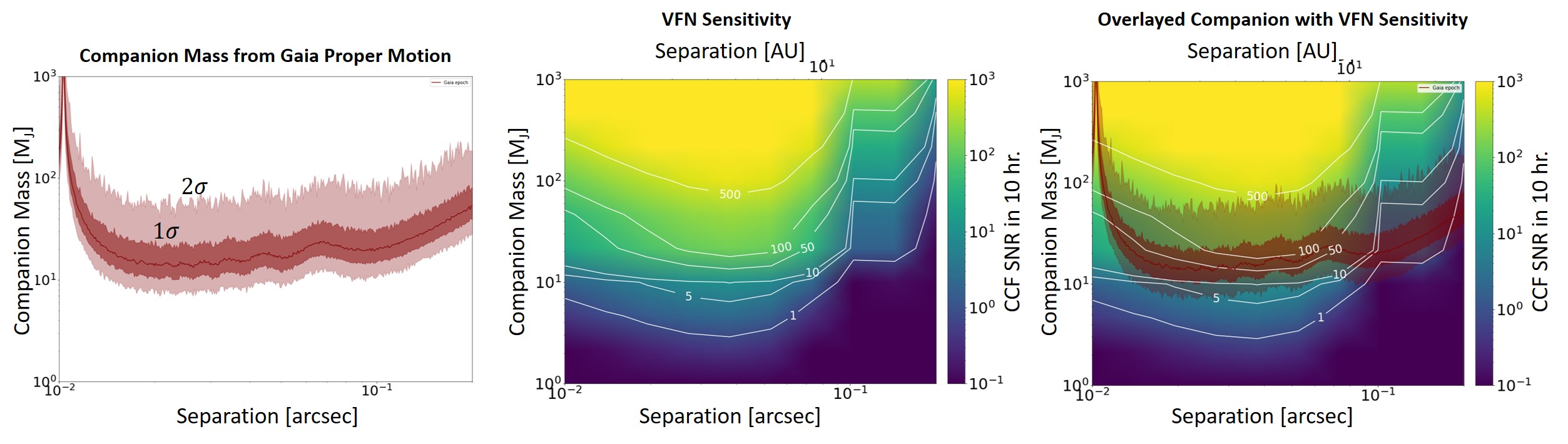}
    \caption{Example of a companion candidate considered for VFN observation. (Left) Companion mass as a function of separation for a sample target from our Gaia-Hipparcos-guided target list. The solid line marks the most probably mass estimate while the darker and lighter shaded regions are 1 and 2$\sigma$ bounds respectively. (Middle) VFN sensitivity to companions around the host star for the companion shown in the plot to the left. The colors denote the CCF~SNR achieved after 10 hours of observation with KPIC VFN and a charge 1 vortex mask. Various contours are shown in white on the plot. (Right) Overlay of the left two plots for easy comparison and determination of the VFN sensitivity to this sample companion candidate.}
    \label{fig:PSISIM_GaiaCompanion}
\end{figure}

Knowing that KPIC VFN is sensitive to many companions around YMG stars, we can start to plan a focused survey guided by astrometry data such as the latest results from the Gaia mission. Here we start with the Hipparcos-Gaia Catalog of Accelerations (HGCA) presented in Brandt 2021\cite{Brandt2021_HipGaiaCatalog}. The HGCA comes from a detailed cross-calibration of Gaia eDR3\cite{Gaia2021_eDR3} and Hipparcos astrometry, and provides statistically robust astrometric accelerations, when present and detected by Gaia or Hipparcos, for almost all the stars that have Hipparcos astrometric solutions. We filter this list for stars that satisfy the following criteria: (1) their Hipparcos-Gaia acceleration is at least $3\sigma$ significant in either the Gaia or Hipparcos epoch, (2) their declination is $>-40^{\circ}$ such that they are observable from the Keck Observatory, (3) their Gaia re-normalised unit weight error (RUWE) parameter\cite{Lindegren2018_RUWE} is $<1.4$, which ensures good astrometric quality (this removes stars that may be in crowded fields or have close spectroscopic binaries), and (4) based on Gaia eDR3 parallaxes, their astrometric acceleration is compatible with being caused by a substellar companion ($<80 M_{Jup}$) between 0.01 and 5 arcseconds from the star. 
We cross-match the resulting target list with the same YMG membership list from the last section so that we are left with a target list of 66 young, bright stars that are likely to have a companion within the KPIC VFN field of view based on their proper motion and positions as measured by both Gaia and Hipparcos. We will perform two additional filtering steps that will help finalize the target list for early VFN science. First, we will check various online catalogs, including the Washington Double Star Catalog\cite{Mason2021_WDS}, to confirm that none of the targets under consideration are known binaries or have known companions that could explain the measured astrometry. Then we will vet each target using the current, traditional vortex coronagraph in NIRC2 at Keck to see if the companion is readily detectable at wider separations for which VFN would not be sensitive. However, these two steps are still underway and for now we are considering all 66 targets from the list generated with the aforementioned procedure.

As explained in Kervella et al. 2019\cite{Kervella2019_GaiaAccel}, the Gaia and Hipparcos astrometry can also be combined to predict the companion mass for a given separation such that the companion would produce the measured proper motion anomaly. The mass and separation are degenerate so this procedure results in a set of mass-separation combinations, as shown in the left plot of Fig.~\ref{fig:PSISIM_GaiaCompanion} for one of our targets. For this calculation, we also consider the unknown orbital elements, including eccentricity and inclination, and randomize over them so as to compute a most-likely mass along with 1 and 2$\sigma$ uncertainties on the mass. This particular target is likely to have a mass of about $11~M_{Jup}$ if it orbits at a separation of between 10 and 100~mas. We determine the KPIC VFN sensitivity to a companion around the given host star using PSISIM by entering the star properties from online catalogs and computing the CCF~SNR at various separations and masses for the companion. The middle plot in Fig.~\ref{fig:PSISIM_GaiaCompanion} shows the results of that simulation, assuming a charge 1 vortex and 10 hours of observation on the target, with yellow representing a strong detection while dark blue represents a poor detection. Looking at the contour corresponding to a CCF~SNR of 5, we find that KPIC VFN would be able to detect a 10 Jupiter mass planet orbiting this star between 10 and 100~mas. Finally, we can overlay the two plots as shown in the right plot from Fig.~\ref{fig:PSISIM_GaiaCompanion} to see how detectable the possible Gaia-Hipparcos candidate would be. Most of the planet mass prediction is above the contour for a CCF~SNR of 10 such that if this candidate is real, KPIC VFN should be able to detect it. 

Figure~\ref{fig:PSISIM_GaiaCompanion} shows just one of many candidates that we are considering for the first survey with KPIC VFN once the mode is commissioned towards the end of next year. This highlights the power of VFN in that it targets planets at small separations where they are easier to detect and are most likely to exist. Several other groups have proposed to do similar surveys but these are mostly limited to searching larger separations due to the inner working angle of their coronagraphs. As such, KPIC VFN is complementary to those surveys since it lets us search closer to the star thereby providing a more complete picture of the orbital space and increasing the likelihood that the possible companion will be detected at all.


\section{SUMMARY}
\label{sec:Summary}
In this paper we reviewed the VFN concept, how it can be implemented in current and planned instruments, and the benefits it carries over conventional coronagraphs. We expanded on previous monochromatic experiments to add laboratory results that demonstrate the VFN concept in broadband light. We presented null depths on the order of $4\times10^{-5}$ in 10\% bandwidth light centered at 650~nm and showed that the null remains under $10^{-4}$ at up to 15\% bandwidth. We also achieved peak planet coupling efficiencies of $\sim8$\% across all bandwidths which is close to the theoretical maximum of 10\% for a charge 2 mask. We confirmed that these results are consistent with those predicted by our system model. At small bandwidths the null is wavefront-error-limited while at larger bandwidths it is limited by chromatic leakage from the vector vortex mask. This experimentally validates the VFN concept at bandwidths commensurate with typical astronomical bands used in exoplanet observations.

We then reviewed the VFN demonstrator mode of the KPIC instrument and provided yield estimates for KPIC VFN. We explained how PSISIM performs a simulation of an exoplanet observation and then provided a framework through which we can test the detectability of hypothetical low-mass companions around young stars given known exoplanet occurrence rates determined from years of RV surveys. This analysis showed that KPIC VFN is sensitive to planets around YMG stars and a blind observation campaign, though inefficient, would yield many low-mass companions including exoplanets down to 2 Jupiter masses. We also provided a method, following previous work by other groups, for finding promising YMG stars based on combined Gaia and Hipparcos data to find anomalous accelerations that could indicate a sub-stellar companion. We presented a simulation of a single object from our most recent target list, to show that KPIC VFN will likely detect that companion assuming it exists. We are finalizing a full target list which we can use for an observation campaign with KPIC VFN towards the end of 2022. Such a campaign could yield several exoplanet detections along with high resolution spectra for those planets and it would be entirely complementary to similar campaigns underway with conventional coronagraph systems.


\acknowledgments 
Daniel Echeverri is supported by a NASA Future Investigators in NASA Earth and Space Science and Technology (FINESST) fellowship under award \#80NSSC19K1423. This work was supported by the Heising-Simons Foundation through grants \#2015-129, \#2017-318, and \#2019-1312. Part of this research was carried out at the Jet Propulsion Laboratory, California Institute of Technology, under a contract with the National Aeronautics and Space Administration (80NM0018D0004). 

The authors wish to recognize and acknowledge the very significant cultural role and reverence that the summit of Maunakea has always had within the indigenous Hawaiian community. We are most fortunate to have the opportunity to conduct observations from this mountain.


\small
\bibliography{Library} 
\bibliographystyle{spiebib} 

\end{document}